\documentclass[nofootinbib,floats,aps,superscriptaddress,preprint]{revtex4} 
 
\newcommand{\beq}{\begin{equation}}
\newcommand{\eeq}{\end{equation}}
\newcommand{\snu}{\tilde \nu}

\newcommand{\mathbold}[1]{\mbox{\boldmath $\bf#1$}}

\newenvironment{Eqnarray}%
         {\arraycolsep 0.14em\begin{eqnarray}}{\end{eqnarray}}
\def\ifmath#1{\relax\ifmmode #1\else $#1$\fi}
\def\ls#1{\ifmath{_{\lower1.5pt\hbox{$\scriptstyle #1$}}}}
\def\beqa{\begin{Eqnarray}}
\def\eeqa{\end{Eqnarray}}
\def\crr{\crcr\noalign{\vskip .15in}}

\def\vev#1{\left\langle #1\right\rangle}

\def\ifmath#1{\relax\ifmmode #1\else $#1$\fi}
\def\eighth{\ifmath{{\textstyle{1 \over 8}}}}

\def\quarter{\ifmath{{\textstyle{1 \over 4}}}}

\def\Ref#1{ref.~\cite{#1}}
\def\Rref#1{Ref.~\cite{#1}}
\def\eq#1{eq.~(\ref{#1})}
\def\Eq#1{Eq.~(\ref{#1})}
\def\eqs#1#2{eqs.~(\ref{#1}) and (\ref{#2})}

\def\cross{\times}
\def\ifmath#1{\relax\ifmmode #1\else $#1$\fi}

\def\sqhalf{\ifmath{{\textstyle{1 \over \sqrt{2}}}}}
\def\eighth{\ifmath{{\textstyle{1 \over 8}}}}

\begin{document}

\preprint{
\vbox{
      \hbox{SCIPP-02/26}
      \hbox{hep-ph/0210273}
      \hbox{October 2002}
    }}
\vspace*{1cm}

\title{The would-be majoron in R-parity-violating supersymmetry} 
\author{Yuval Grossman}  
\affiliation{Department of Physics,  
Technion--Israel Institute of Technology,\\  
       Technion City, 32000 Haifa, Israel\vspace{6pt}}  

\author{Howard E. Haber}  
\affiliation{Santa Cruz Institute for Particle Physics \\
University of California, Santa Cruz, CA 95064 USA
\\[20pt] $\phantom{}$ }

\begin{abstract}%
In lepton-number-violating supersymmetric models, there is no natural
choice of basis to distinguish the down-type Higgs and lepton
superfields.  We employ basis-independent techniques to identify the 
massless majoron and associated light scalar in the case of
spontaneously-broken lepton number (L).  When explicit L-violation is added,
these two scalars can acquire masses of order the electroweak scale
and can be identified as massive sneutrinos.

\end{abstract}

\maketitle

\section{Introduction}

Recent data that exhibits neutrino mixing phenomena implies that
the lepton sector of the Standard Model must be extended \cite{review}.
The simplest extension involves adding right handed neutrinos,
and then tuning the neutrino masses to be less than $\mathcal{O}(1~{\rm
eV})$ (if neutrinos are Dirac fermions) or by invoking the seesaw
mechanism (if neutrinos are Majorana fermions). 
In low-energy supersymmetric models, it is possible to introduce
neutrino masses in a phenomenologically acceptable way
without adding right-handed neutrinos.  One simply
allows for renormalizable terms that violate 
lepton number (L), while imposing baryon number (B)
invariance.  This can be achieved by replacing R-parity of the minimal
supersymmetric model (MSSM) with a ${\bf
Z_3}$ triality~\cite{triality}.  
This model provides an alternative framework for
neutrino masses.  Eventually, one must try to understand why the
L-violating parameters of the model are small enough to yield neutrino
masses at the observed level~\cite{Banks}. 

In the B-conserving, L-violating alternative to the MSSM, the
L-violating terms are explicit.  One can also generate L-violation
directly in the MSSM if one of the sneutrinos acquires a vacuum
expectation value~\cite{susynu}.  
In the latter case, L is spontaneously broken,
which implies that a massless Goldstone boson, the majoron, must exist
in the spectrum~\cite{originalmajoron}.  
Since the sneutrino is an electroweak doublet, one
can show that the spectrum must also include a very light CP-even
scalar partner to the CP-odd majoron~\cite{lightscalar}.\footnote{This
is a feature of both the non-supersymmetric and supersymmetric doublet
majoron models.}
Models of this type are excluded
since the decay of the $Z$ into the majoron and its CP-even scalar
partner is not observed~\cite{susymajoron,lightscalar}.  
Thus, any viable L-violating supersymmetric model whose 
field content is identical to that of the MSSM must possess explicit
L-violating terms.
There are also ways to extend the model of
spontaneous L-violating supersymmetry by adding additional chiral superfields
(including electroweak singlets) such that the majoron is dominantly a
singlet and all other scalar masses lie above $m_Z$~\cite{sing-maj}.  
However, such models lie outside the scope of this paper.

We consider the most general L-violating low-energy supersymmetric
model, with the MSSM field content.  In addition to the 
effects of the explicit L-violating terms, one must also consider
the L-violating effect that depends on the vacuum expectation values
of the sneutrino fields.  Of course, the latter is
basis-dependent, and it is often convenient to define the Higgs field
such that the orthogonal physical sneutrino fields have no vacuum
expectation value.  However, other choices are possible, which suggests
that the model can be viewed as a model of spontaneously-broken lepton
number with additional explicit L-violating terms.  Since models of
spontaneously-broken lepton number possess a massless majoron, when
explicit L-violating terms are included, the majoron acquires a
squared-mass proportional to the relevant explicit 
lepton-number-violating term.
Two questions immediately arise: (i)~how do we identify the
would-be majoron? and (ii)~if explicit lepton-number violation is very
small (which is needed to explain the magnitude of neutrino masses), 
how does one avoid a very light would-be majoron?  These questions
have been previously examined in the literature~\cite{gato,comelli}. 
In this short note, we revisit both these questions and demonstrate
how they can be addressed in a basis-independent formalism~\cite{Dav,ghbasis}.

\section{The scalar potential and minimum conditions}

In the notation of \Ref{GHrpv},
the contribution of the neutral scalar fields to the
scalar potential, before imposing L-conservation, is
\beqa \label{Vn}
V_{\rm neutral} &=& \left(m_U^2+ |\mu|^2\right) {|h_U|}^2 +
\left[({M^2_{\tilde L}})_{\alpha\beta} + \mu_\alpha\mu_\beta^*\right]
\snu_\alpha \snu_\beta^{*} - \left(b_\alpha\snu_\alpha h_U +b_\alpha^*
\snu_\alpha^* h_U^*\right) \\ &&\quad +\eighth(g^2+g^{\prime 2})
\left[|h_U|^2-|\snu_\alpha|^2 \right]^2\,, \nonumber 
\eeqa 
where $h_U$ is the neutral component of the up-type scalar doublet,
and we have combined the neutral component of the down-type scalar
doublet, $\snu_0\equiv h_D$ and the three sneutrinos, $\snu_i$ into a
generalized sneutrino field $\snu_\alpha$, where $\alpha=0, \cdots,
n_g$ (for $n_g=3$ generations). 
In minimizing the full scalar potential, we assume that only neutral
scalar fields acquire vacuum expectation values: $\langle
h_U\rangle\equiv \sqhalf v_u$ and $\langle\snu_\alpha\rangle\equiv 
\sqhalf v_\alpha$.  From \eq{Vn}, the minimization conditions are:
\beqa 
(m_U^2+|\mu|^2)v_u^* &=& b_\alpha v_\alpha-\eighth(g^2+g^{\prime 2})
(|v_u|^2-|v_d|^2)v_u^*
\,,\label{mincondsa} \\
\left[({M^2_{\tilde L}})_{\alpha\beta} + \mu_\alpha\mu_\beta^*\right]
v_\beta^* &=& b_\alpha
v_u+\eighth(g^2+g^{\prime 2})(|v_u|^2-|v_d|^2)v_\alpha^*\,,\label{mincondsb} 
\eeqa
where   
\beq\label{vdef}
|v_d|^2\equiv \sum_\alpha |v_\alpha|^2\,.
\eeq
The normalization of the vacuum expectation values has been chosen
such that
\beq \label{vevdef}
v\equiv (|v_u|^2+|v_d|^2)^{1/2}={2m_W\over g}=246~{\rm GeV}\,.
\eeq

It is convenient to introduce two additional quantities.  We define:
\beq
\label{treelevelsnumasses}
M^2_{\alpha\beta}\equiv
({M^2_{\tilde L}})_{\alpha\beta}+ \mu_\alpha\mu_\beta^\ast-\eighth 
(g^2+g^{\prime 2})(v_u^2-v_d^2)\delta_{\alpha\beta}\,.
\eeq
Using this quantity, we can simplify the second minimum condition
[\eq{mincondsb}] which now reads\footnote{Note that one can always choose the 
vacuum expectation values $v_u$ and $v_\alpha$ 
real by suitable phase re-definitions of the scalar fields.
Henceforth, we assume that all vacuum expectation values are taken 
to be real.}
\beq \label{mincondition}
M^2_{\alpha\beta}v_\beta^\ast=v_u b_\alpha\,.
\eeq
It is also useful to define the vector $c_\alpha$ as follows
\beq \label{cdef}
M^2_{\alpha\beta}b_\beta=|b|^2 c_\alpha\,,
\eeq
where $|b|^2\equiv\sum_\alpha b_\alpha^* b_\alpha$.

\section{Spontaneous Lepton Number Violation in the MSSM}

We begin by considering the possibility of spontaneous L-violation in
low-energy R-parity-conserving (RPC) supersymmetry consisting only of the
MSSM fields.  We impose L-conservation on the MSSM Lagrangian, which
constrains the scalar potential [\eq{Vn}].  In the usual
basis choice in which $h_D$ is a Higgs field and $\snu_j$ ($j=1,2,3$)
are the lepton number carrying sneutrino fields, it follows 
that $\mu_\alpha=(\mu, 0,0,0)$,
$b_\alpha=(b, 0,0,0)$, and $({M^2_{\tilde L}})\ls{j 0}=
({M^2_{\tilde L}})\ls{0 j}=0$.
Note that $\snu_0\equiv h_D$ and the $\snu_j$ transform the same way
under the SU(3)$\times$SU(2)$\times$U(1) gauge symmetry, but are
distinguished by their L quantum numbers: $\snu_0$ is neutral while
the $\snu_j$ possess non-zero L.  Hence if any of the $\snu_j$ acquire
a vacuum expectation values, L will be spontaneously 
broken.  Noting that $M^2_{j0}=M^2_{0j}=0$ in the basis defined above,
\eq{mincondition} implies that $M^2_{ij}v_j=0$. Thus, if at least 
one of the
$v_j$ is non-zero, it follows that ${\rm det}~M^2=0$.  This is a
necessary (basis-independent) 
condition for spontaneous lepton number violation.
%Moreover, this condition is basis-independent since the
%determinant is invariant under change of basis.

We assume that $v_u\neq 0$ and $v_0\neq 0$.\footnote{If 
$v_1=0$, then \eqs{mincondsa}{mincondsb} simply
reduce to the usual RPC MSSM equations for $v_u$ and $v_d=v_0$.  
If we had assumed that $v_u=v_0=0$, 
%(note that $v_u=0$ implies that $v_0=0$ and vice versa), 
then one finds that lepton number is spontaneously
broken with $v_1^2= -8({M^2_{\tilde L}})_{11}/(g^2+g^{\prime
2})$.  In this case, there is
a consistent solution if $({M^2_{\tilde L}})_{11}<0$.
However, a model with $v_u=v_0=0$ would not generate any quark masses,
so we will not consider this case any further.}    
Without loss
of generality, we may perform a rotation of the sneutrino fields among
the $\snu_j$ such that $v_1\neq 0$ while $v_2=v_3=0$.
It then follows from \eqs{mincondsa}{mincondsb} that:
\beqa
&& ({M^2_{\tilde L}})_{11}=\eighth (g^2+g^{\prime
2})(v_u^2-v_0^2-v_1^2)\,, \\[5pt]\label{spla}
&& (m_U^2+({M^2_{\tilde L}})_{11}+|\mu|^2)v_u=bv_0\,, \\[5pt]\label{splb}
&& (m_D^2-({M^2_{\tilde L}})_{11}+|\mu|^2)v_0=bv_u\,,\label{splc}
\eeqa
where $m_D^2\equiv ({M^2_{\tilde L}})_{00}$.
These equations have a consistent solution for non-zero $v_u$, $v_0$
and $v_1$ only if 
\beq \label{detcond}
(m_U^2+({M^2_{\tilde L}})_{11}+|\mu|^2)
(m_D^2-({M^2_{\tilde L}})_{11}+|\mu|^2)=b^2\,.
\eeq
For this very particular choice of parameters, the quantities
$v_u/v_0$ and $v_u^2-v_0^2-v_1^2$ are fixed, but this is not enough
information to determine all three vacuum expectation values 
uniquely at tree-level.  That is, there is a flat direction 
in the scalar potential at tree-level.
%Note that \eq{splc} is equivalent to the condition: $M^2_{00}v_0= v_u
%b$ [see \eq{condition}]. 
\Rref{gato} demonstrates that by considering the renormalization group
evolution of the potential parameters, 
there is generically some momentum scale $Q_0$ for which \eq{detcond} 
is satisfied.  Then, when the one-loop effective potential is
evaluated, the flat direction is lifted and the undetermined vacuum
expectation value is fixed via dimensional transmutation in terms of $Q_0$.
The parameters of the model must be tuned to get the
observed  $Z$ mass,
$m_Z^2=\quarter(g^2+g^{\prime 2})(v_u^2+v_0^2+v_1^2)$, as well as the
correct hierarchy $v_1\ll v$ needed to explain the light neutrino mass.
 
If lepton number is spontaneously broken, then there must be a
massless Goldstone boson---the majoron~\cite{originalmajoron}.  
We shall exhibit this explicitly
in the case above where \eq{detcond} holds.   For simplicity, we 
assume that the model is CP-conserving.\footnote{In a basis where the vacuum
expectation values are real, it then follows that $\mu_\alpha$, $b_\alpha$
and $({M^2_{\tilde L}})_{\alpha\beta}$ are real.}   We can then compute
CP-even and CP-odd scalar squared-mass matrices.  In \Ref{ghbasis}, we
showed that after removing the Goldstone boson that gives mass to the $Z$,
the CP-odd scalar squared-mass matrix in a general L-violating model 
%(where L need not be a good symmetry) 
is given by 
\beq \label{modd2d2}
M_{\rm odd}^2= \pmatrix{
v^2 (v \cdot b)/(v_u v_d^2) & 
v b_\beta X_{\beta i}/v_d \cr
v X_{j\alpha}b_\alpha/v_d & 
X_{j \alpha} M^2_{\alpha\beta} X_{\beta i}
}\,, 
\eeq
where $v\cdot b\equiv v_\alpha b_\alpha$
%$v_d$, $v^2$ and $M^2_{\alpha\beta}$ are defined in
%\eqs{vdef}{treelevelsnumasses},
and the $X_{\beta i}$ are chosen
so that the set $\{v_\beta/v_d,X_{\beta i}\}$
forms an orthonormal set of vectors in an $(n_g+1)$-dimensional vector
space (for $n_g$ generations).  
In our notation, $X_{j\alpha}\equiv X^T_{\alpha j}$, where the
superscript $T$ denotes the matrix transpose.
The following relations will be useful:
%\beqa 
%v_\alpha X_{\alpha i} &=& 0\,, \label{ort1} \\
%X_{\alpha i}X_{\alpha j} &=& \delta_{ij}\,, \label{ort2} \\
%X_{\alpha i}X_{\beta i} &=& \delta_{\alpha \beta} 
%   - {v_\alpha v_\beta \over v_d^2}\,. \label{ort3} 
%\eeqa
\beq 
v_\alpha X_{\alpha i} = 0\,, \qquad \label{ort1} 
X_{\alpha i}X_{\alpha j} = \delta_{ij}\,, \qquad \label{ort2} 
X_{\alpha i}X_{\beta i} = \delta_{\alpha \beta} 
   - {v_\alpha v_\beta \over v_d^2}\,. \label{ort3} 
\eeq
To show that there is a majoron in the case of spontaneously broken 
L, we exhibit the eigenvector of $M_{\rm odd}^2$ with zero
eigenvalue.  Consider the eigenvector:\footnote{Although the cross
product technically exists only in three dimensions, the dot product
of two cross products can be expressed in terms of dot products and
thus exists in any number of dimensions.  For example,
$|v\times b|^2= v_d^2 b^2-(v\cdot b)^2$.  Note that by assumption in
this calculation, $b_\alpha=(b,0,0,0)$ and $v_\alpha=(v_0,v_1,0,0)$
with $v_1\neq 0$.  Hence $|v\times b|^2\neq 0$.}
\vspace*{0.12in}
\beq \label{majoron}
J_\beta\equiv
\pmatrix{\displaystyle\frac{-v_u}{v} \crr \displaystyle
\frac{v_d (v\cdot b) b_\rho X_{\rho i}}{|v\times b|^2}}\,,
\eeq

\vspace*{0.12in}
\noindent
A simple calculation shows that $(M_{\rm odd}^2)_{\alpha\beta}J_\beta=0$
[after applying \eq{mincondition}], if
\beq \label{majcondition}
(v\cdot b)M^2_{\alpha\beta} b_\beta-v_u b^2 b_\alpha=0\,.
\eeq
It is easy to check that \eq{majcondition} is satisfied under the
assumption of L conservation of the MSSM
Lagrangian.\footnote{L conservation implies that one can choose a 
basis in which $M^2_{0j}=M^2_{j0}=b_j=0$.
\Eq{mincondition} then implies that $M_{00}^2=v_u b/v_0$.}  
It is interesting to note that \eq{majcondition} can be written more
simply as $b_\alpha= (v\cdot b/v_u)c_\alpha$, where $c_\alpha$ is defined in
\eq{cdef}.  It then follows that $|b\times c|^2=0$, and we conclude that the
necessary and sufficient basis independent
condition for spontaneously broken lepton number is $|b\times c|^2=0$,
with $|v\times b|^2\neq 0$.\footnote{Note that from \eq{mincondition},
$|v\times b|^2=0$ implies that $|b\times c|^2=0$, but the converse is
true only if $M^2_{\alpha\beta}$ is an invertible matrix.  But, we
noted previously that ${\rm det}~M^2=0$ is a necessary condition for 
 spontaneously broken lepton number.}

We now turn to the CP-even scalar that is associated with the CP-odd
majoron.  Again following \Ref{ghbasis}, the CP-even
scalar squared-masses of the model can be determined by computing the
eigenvalues of the following squared-mass matrix:
\beq \label{meven2d2}
M_{\rm even}^2= 
 \pmatrix{m_Z^2\cos^2 2\beta &  -m_Z^2\sin 2\beta\cos 2\beta & 0 \cr
 -m_Z^2\sin 2\beta\cos 2\beta & 
m_Z^2\sin^2 2\beta+v^2 (v \cdot b)/(v_u v_d^2) & 
-v b_\beta X_{\beta i}/v_d \cr
0 & -v X_{j\alpha}b_\alpha/v_d & 
X_{j \alpha} M^2_{\alpha\beta} X_{\beta i}
}\,, 
\eeq
where $\tan\beta\equiv v_u/v_d$, with $v_d$ given by \eq{vdef}.  
First, we note that if $\cos 2\beta=0$, then there is a massless
scalar state at tree-level in all circumstances ({\it i.e.}, conserved
L, spontaneously broken L or explicitly broken L).  In the case of
spontaneously broken L, we can identify this state as the massless 
scalar state associated with the majoron.  Henceforth, we shall assume
that $\cos 2\beta\neq 0$. 
%(or more precisely, $|\cos 2\beta|\gg v_L/v$).
Then, one can easily verify that the eigenvector
\vspace*{0.12in}
\beq \label{cpevenrho}
\rho_\beta\equiv
\pmatrix{\displaystyle\frac{v_u\sin 2\beta}{v\cos 2\beta} \crr
\displaystyle\frac{v_u}{v} \crr \displaystyle
\frac{v_d (v\cdot b) b_\rho X_{\rho i}}{|v\times b|^2}}\,,
\eeq

\vspace*{0.12in}
\noindent
satisfies $(M_{\rm even}^2)_{\alpha\beta}\rho_\beta=0$ provided that
\eq{majcondition} holds.  That is, there exists a massless CP-even
scalar at tree-level, $\rho$, associated with the massless majoron,
$J$.  When radiative corrections are incorporated, the mass of $\rho$
is not protected (it is not a Goldstone boson).  Thus, $\rho$ gains a
small mass of $\mathcal{O}(v_1)$.  Nevertheless, the experimental
absence of the decay $Z\to J\rho$ implies that the model of
spontaneously broken R-parity described above is ruled out.

\section{Explicit L-violation and the would-be majoron}
  
We now consider the introduction of explicit L-violating terms.
Clearly, the majoron eigenstate identified in \eq{majoron} is no longer an
eigenstate of the CP-odd squared-mass matrix.  But, to the extent that
explicit L-violation is small, the majoron identified above is an
approximate eigenstate, but with a non-zero mass.  We denote this
state as the would-be majoron.  It is a simple matter to
use first-order perturbation theory to compute its mass.  

Suppose we write: $M^2_{\rm odd}= M^{(0)2}_{\rm odd}+M^{(1)2}_{\rm
odd}$, where $J_\beta$ [\eq{majoron}] is the eigenvector of
$M^{(0)2}_{\rm odd}$ with zero eigenvalue.  
Using first order perturbation theory, the squared-mass is computed by
evaluating the expectation value of $M^{(1)2}_{\rm odd}$ with respect
to the unperturbed normalized eigenvalue ({\it i.e.}, $J_\beta$
normalized to unit length).   Since the unperturbed majoron is massless,
this is equivalent to computing the expectation value of the full
squared-mass matrix $M^2_{\rm odd}$.  Thus, the squared-mass of the
would-be majoron, $J$, is
\beq
m^2_J= {(M^2_{\rm odd})_{\alpha\beta}J_\alpha J_\beta \over N_o^2}\,,
\eeq
where $N_o^2\equiv \sum_\alpha J_\alpha J_\alpha$.  After much algebraic
simplification, the end result is:
\beq \label{wouldbee}
m^2_J={v_d^2 v^2(v\cdot b)\left[(v\cdot
b)M^2_{\alpha\beta}b_\alpha b_\beta-v_u b^4\right]\over
|v\times b|^2\left[v_u^2\,|v\times b|^2+v^2(v\cdot b)^2\right]}\,,
\eeq
where $b^2\equiv\sum_\alpha b_\alpha b_\alpha$.  It is useful to
define the basis-independent quantity:
\beq \label{vldef}
v_L^2\equiv {|v\cross b|^2\over b^2}=v_d^2-{(v\cdot b)^2\over b^2}\,.
\eeq
Note that in a basis where $b_j=0$, one obtains 
$v_L^2\equiv v_d^2-v_0^2=\sum_i v_i^2$.  That is, $v_L\ll v_d$, 
assuming that L-violating effects are small. Hence, we can drop
%$|v\times b|^2\ll v^2(v\cdot b)^2/v_u^2$, and we can therefore drop
the first term relative to the second in the denominator of \eq{wouldbee}.
In addition, using the definition of
$c_\alpha$ [\eq{cdef}], the above result can be further simplified.
We then obtain:
\beq \label{wouldbe}
m^2_J={v_d^2 (v\times b)\cdot(b\times c)\over (v\cdot b)v_L^2}\,.
\eeq
Note that if we go to the spontaneous L-violating limit in which 
$|b\times c|=0$ [with $v_L\neq 0$], one finds a massless majoron
as expected.  Further, in the case of explicit L violation, it is easy
to check that 
$v_L\neq 0$.\footnote{In a basis where $b_i=0$, $v_L=0$
implies that $v_i=0$.  Then from \eq{mincondition} 
one obtains $M^2_{i0}=0$.  
In this case, barring the unlikely cancellation 
$M^2_{i0}=({M^2_{\tilde L}})_{i0}+ \mu_i\mu_0=0$ for non-vanishing 
$({M^2_{\tilde L}})_{i0}$ and $\mu_i$, it follows that
the scalar potential is L-conserving in
contradiction to our assumption.}  One notable feature of
\eq{wouldbe} is that it provides a basis-independent expression for
the mass of the would-be majoron.

Finally, we can address the puzzle of how the would-be majoron mass
can be of ${\mathcal{O}}(m_Z)$ even if the explicit L-violation is
small~\cite{comelli}.  
It is convenient to choose a basis in which $b_i=0$.  Using
the minimum condition [\eq{mincondition}] and \eq{wouldbee},
and assuming that $v_L\ll v$,
%the explicit L-violation is small,
%where $v_L^2\equiv v_d^2-v_0^2=\sum_i v_i^2$.  We have
%By assumption, the explicit L-violation is small.  In the 
%(approximately) lepton number conserving basis, this
%is equivalent to $v_L\ll v_0$, and 
we end up with:
\beq \label{jassnu}
m_J^2={\sum_{ij} M_{ij}^2 v_i v_j\over \sum_i v_i^2}
\biggl[1+\mathcal{O}\left({v_L^2\over v^2}\right)\biggr]\,.
\eeq
To understand the physical implication of this result, let us choose
the direction of $v_i$ to point along the $k$th direction.  Then
$m_J^2=M_{kk}^2$.  But, in the limit of small explicit
L-violation, $M_{kk}^2$ is the squared-mass of the $k$th sneutrino
(in the RPC limit).
Thus we have identified the would-be majoron as one of the
sneutrinos.  Since the model parameters can easily be chosen such that
$M_{kk}^2\sim {\mathcal{O}}(v^2)$, we see that there is no contradiction
in having the would-be majoron mass of ${\mathcal{O}}(v)$, even in the
limit of arbitrarily small explicit L breaking~\cite{comelli}.
Nevertheless, the limit of vanishing explicit lepton number violation
is smooth.  In particular, note that for $b_i=0$, 
\eq{mincondition} implies that
$M_{i0}^2v_0= -M_{ij}^2 v_j$.  In the limit of an L-conserving
Lagrangian in which L is spontaneously broken, $M_{i0}=0$ while one of
the $v_i$ is nonzero.  This implies that $M_{ij}^2 v_i v_j=0$ and the
massless majoron is regained.  

These results can also be understood in
a basis-independent language using the results of \eq{wouldbe}.  The
squared-mass 
of the would-be majoron is proportional to the dimensionless ratio 
of two small parameters, $(v\cross b)\cdot(b\cross c)/[(v\cdot b) v_L^2]$.
The numerator is a consequence of explicit L-breaking and the
denominator is proportional to the square of the sneutrino vacuum
expectation value in the case of spontaneous L-breaking.
Nevertheless, the ratio of these two small quantities
can be $\mathcal{O}(1)$, in which case $m_J$
is of order the electroweak scale.

To see that this last result does not contradict our usual intuition
about explicit symmetry breaking, consider for simplicity the one
generation case.  Then, we can write
$m_J^2=M_{11}^2=-M_{10}^2 v_0/v_1$.  We then see explicitly
that $m_J^2$ is {\it linear} in the explicit L-violating
parameter $M_{10}^2$.  Nevertheless, in the limit of small $M_{10}^2$,
because $M_{10}^2/v_1$ can be of the same order as $v_0$, it follows
that $m_J^2$ can be of ${\mathcal{O}}(v^2)$ without an unnatural
tuning of the parameters.
%\footnote{Roughly speaking, the squared-mass
%of the would-be majoron is the ratio of an explicit R-parity-violating
%parameter to the scale of spontaneous R-parity violation.  Since
%phenomenology requires the latter to be very small [rather than of
%order the electroweak symmetry breaking scale], it is possible to have
%the ratio of these two small parameters {\it unsuppressed} 
%as compared to the scale of electroweak symmetry breaking.}
A simple exercise shows that this is in
accord with the expectations of Dashen's formula~\cite{dashen}.  
For example, consider
the linear O(4) sigma model~\cite{chiral} consisting of $\sigma$ and
${\mathbold{\vec\pi}}$, with the usual Mexican hat potential and
corresponding vacuum expectation value $v$.  If we now
break the O(4) symmetry with ${\mathcal{L}}_{\rm break}
=a\sigma$, then the Goldstone boson ($\pi$)
acquires a mass that is linear in $a$ and is given by Dashen's formula:
\beq
v^2 m_\pi^2 = \vev{0|[Q,[Q,{\mathcal{L}}_{\rm break}]]|0}
= av\,,
\eeq
where $Q$ is the Noether symmetry charge and $v=\vev{0|\sigma|0}$ is
the vacuum expectation value in the absence of explicit symmetry 
breaking.  Thus, $m_\pi^2=a/v$, which has the same behavior as 
$m_J^2 \propto M_{10}^2/v_1$.  Of course, in QCD the relevant
chiral symmetry breaking
parameters are such that $m_\pi\ll\Lambda\sim 4\pi v$~\cite{georgi}.
In contrast, one must choose $M_{10}^2\sim{\mathcal{O}}(v_0 v_1)$
in order to ensure that the sneutrino mass is of order the electroweak
scale (light sneutrinos are ruled out
by the absence of $Z$ decay into sneutrino pairs).
%The main difference between the
%two cases is that in the explicitly broken linear sigma model, 
%$v$ is independent of the explicit symmetry breaking parameter $a$.
%In our case, $v_1$ and $M_{10}^2$ are related in the $b_i=0$ basis.

For completeness, we evaluate the mass of the CP-even scalar $\rho$
associated with the majoron when explicit L violation is
introduced.  Following the method of computation of $m_J^2$, we again
use first-order perturbation theory.  Writing 
$M^2_{\rm even}= M^{(0)2}_{\rm even}+M^{(1)2}_{\rm
even}$, and using the fact that $\rho_\beta$ [\eq{cpevenrho}] 
is an eigenvector of $M^{(0)2}_{\rm even}$ with zero eigenvalue, 
it follows that
\beq
m_\rho^2={(M^{2}_{\rm even})_{\alpha\beta}\rho_\alpha\rho_\beta\over N_e^2}\,,
\eeq
where $N_e^2\equiv\sum_\alpha \rho_\alpha\rho_\alpha$.  The end result
is:
\beq \label{rhomass1}
m^2_\rho={v_d^2 v^2 (v\cdot b)(v\times b)
\left[(v\cdot b)M^2_{\alpha\beta}b_\alpha b_\beta-v_u b^4\right]
\cos^2 2\beta\over |v\times b|^2\left[v_u^2\,|v\times b|^2
+v^2(v\cdot b)^2\cos^2 2\beta\right]}\,.
\eeq
As noted previously [see discussion below \eq{meven2d2}], if 
$\cos 2\beta=0$, then $m_\rho=0$ is an exact tree-level result, even
in the presence of L-violating terms.  Assuming that $|\cos 2\beta|\gg
v_L/v$ and that L-violating effects are small, we may
again drop the first term relative to the second in the denominator of
\eq{rhomass1}.  As before, we  obtain:
\beq \label{rhomass2}
m^2_\rho={v_d^2 (v\times b)\cdot(b\times c)\over (v\cdot b)v_L^2}\,.
\eeq
That is, 
\beq\label{rhomass3}
m^2_\rho=m^2_J\,\biggl[1+\mathcal{O}\left({v_L^2\over v^2}\right)\biggr]\,.
\eeq
Following the discussion below \eq{jassnu}, we identify $\rho$
as a sneutrino (in the RPC limit).  Moreover, since $\rho$ and $J$ are 
degenerate in the RPC limit, these two real scalars
can be combined to make a (complex) sneutrino
state of definite lepton number~\cite{ghprl}.  

At tree-level,
the squared-mass splitting, $\Delta m^2\equiv m_\rho^2-m_J^2$ is non-zero
when explicit L-violation is present.  The analysis above seems to imply that
$\Delta m^2\sim\mathcal {O}(v_L^2/v^2)$.  However, an explicit
expression for $\Delta m^2$ to first order in $v_L^2/v^2$ would
require a second-order perturbation theory computation of 
$m_J^2$ and $m_\rho^2$.  In the presence of explicit L-violation, if
$m_J$, $m_\rho\sim \mathcal{O}(v)$ then we may use the results 
of \Ref{ghbasis} to
obtain a basis-independent expression for $\Delta m^2$.  
This case
corresponds to sneutrino masses of order the electroweak scale, and we
indeed verify that $\Delta m^2\sim\mathcal {O}(v_L^2/v^2)$.
On the other hand, if $m_J$, $m_\rho\ll v$, then  
the results of \Ref{ghbasis} do not directly apply, since there is an
independent small parameter which must be treated consistently in the
expansion around the L-conserving limit.  In this case, the tree-level
value of
$\Delta m^2$ can be significantly smaller than $\mathcal {O}(v_L^2/v^2)$.
Consequently, one must not neglect the
%Of course, one must then include 
radiative corrections that could end
up as the dominant contribution to the squared-mass difference.

\section{Conclusions}

In models of R-parity-violating supersymmetry, there is no longer a
distinction between the hypercharge $Y=-1$ Higgs superfield and the
lepton superfields.  In computing physical quantities involving the
scalar Higgs and slepton sectors, one can
either choose a basis in the generalized Higgs-lepton flavor space or
employ basis-independent techniques.  For example, one could choose to
define the Higgs field direction so that the neutral slepton vacuum
expectation values vanish.  However, in this case, the distinction
between spontaneous lepton number violation (typically associated with 
non-zero sneutrino vacuum expectation values) and explicit lepton number
violation is unclear.  By employing basis-independent methods, we are
able to provide an unambiguous condition for the existence of
spontaneous lepton number violation.

In the latter case, the spectrum contains a massless Goldstone
boson---the CP-odd majoron.  The simplest models of this type also predict
the existence of a very light CP-even scalar partner.  Such
models are ruled out by precision $Z$ decay data.  Thus, any realistic
L-violating model (based solely on the superfields of the MSSM)
must contain some explicit L-breaking.  The would-be majoron
acquires a squared-mass that depends linearly on the explicit
L-breaking squared-mass parameter.  We demonstrate how to
compute the mass of the would-be majoron using basis-independent
techniques, and identify this CP-odd scalar and its CP-even scalar partner
as approximate sneutrino states.  
Finally, we have shown how it is possible for the mass of the would-be
majoron and its CP-even scalar partner
to be of $\mathcal{O}(v)$ despite the fact that the explicit
L-violation must be small enough to account for neutrino masses
less than of $\mathcal{O}(\rm {eV})$.

%%%%%%%%%%%%%%%%%%%%%%%%%%%%%%%%%%%%%%%%%%%%%%%%%%%%%%%%  

\acknowledgments  

We thank Sacha Davidson for useful discussions.
H.E.H.~is grateful for the hospitality and
support of Oxford University, where this work was initiated.  In
addition, we acknowledge the Aspen Center for Physics and the Stanford
Linear Accelerator Center, where this work was subsequently carried
out.  Y.G.~was supported in part by the Israel Science Foundation under Grant 
No.~237/01 and in part by the United 
States--Israel Binational Science Foundation (BSF) through Grant No. 2000133. 
H.E.H.~is supported in part by the U.S.~Department of Energy
under grant no.~DE-FG03-92ER40689.
  
\clearpage

%%%%%%%%%%%%%%%%%%%%%%%%%%%%%%%%%
%%%%%%%%%%%%%  Ref  %%%%%%%%%%%%%
%%%%%%%%%%%%%%%%%%%%%%%%%%%%%%%%%

\end{document}